\documentclass[amsmath,amssymb,pre,12pt,tightenlines]{revtex4}

\usepackage{graphicx}
\usepackage{bm}

\begin{document}


\title{The conundrum of functional brain networks: small-world efficiency or fractal modularity}
\author{Lazaros K. Gallos$^1$,  Mariano Sigman$^2$, Hern\'an A. Makse$^{1,2}$}

\affiliation{ $^1$ Levich Institute and Physics Department, City
College of New York, New York, New York 10031, USA\\ $^2$ Integrative
Neuroscience Laboratory, Physics Department, FCEyN, Universidad de
Buenos Aires, Buenos Aires, Argentina}

\date{\today}
\begin{abstract}

The human brain has been studied at multiple scales, from neurons, circuits,
areas with well defined anatomical and functional boundaries, to large-scale
functional networks which mediate coherent cognition. In a recent work,
we addressed the problem of the hierarchical organization in the brain
through network analysis. Our analysis identified functional brain
modules of fractal structure that were inter-connected in a small-world
topology. Here, we provide more details on the use of
network science tools to elaborate on this behavior.
We indicate the importance of using percolation theory to highlight
the modular character of the functional brain network.
These modules present a fractal, self-similar topology, identified through
fractal network methods.
When we lower the threshold of correlations to include weaker ties,
the network as a whole assumes a small-world character.
These weak ties are organized precisely as predicted by theory maximizing information transfer with
minimal wiring costs.
\end{abstract}

\maketitle

\section{Introduction}

The functional magnetic resonance imaging (fMRI) technique is a tool that has greatly
improved our ability to probe brain activity. The method detects changes in blood oxygenation
when areas of the brain are activated and consequently require increased blood flow.
In this way, we can monitor what brain areas respond to different mental activities.
The resulting datasets offer a three-dimensional image of the brain indicating the
level of activation at various regimes.

Many methods have been applied to analyze fMRI data, ranging from statistics
to signal processing techniques. Recently, the brain organization has been described
as a complex network \cite{eguiluz2005scale,sporns2005human,bullmore2009complex}.
This approach can take various forms, such as physical connections
between neurons or correlations in the activity between brain areas at a coarser level.
In a recent work \cite{newpnas} we used recent advances in fractal network theory
to characterize the brain clusters structure, and studied one key problem of neuroscience,
namely the integration of modular clusters in a larger scale. Here, we expand on those findings
and describe the methodology in detail, focusing on the use of network theory in the
study of fMRI data.

One of the main features of our sensations is its unitary nature.
The brain can receive many concurrent stimuli. These have to be processed independently
of each other, but at the same time they have to be integrated into a unified entity.
This suggests that the modalities in the brain that process different characteristics
have to act isolated for efficient computations, but they need also be sufficiently
connected in order to perform coherent functions.

The notion of a complex network can be suitably adapted to address this scaling problem
and study optimal information flow in modular networks.
This representation of complicated interactions
has offered new insight in many processes across different disciplines.
A key feature of many such networks is their modular character, a topic which
has attracted a lot of interest in the literature. Many algorithms have
been proposed for the detection of modules, loosely defined as network areas
well-connected within themselves but sparsely connected to the rest of the network.
The detection and behavior of modules at different observation scales, though, remains
a largely unexplored problem.
Network analysis of functional \cite{eguiluz2005scale} and structural \cite{sporns2005human} data has
been used to characterize global connectivity and topological
organization of the human brain\cite{bullmore2009complex}. 
Many of those studies indicate the small-world character \cite{watts98} of brain networks, but
the idea of a simple small-world structure can be contradictory to modular network.

In the present manuscript we implement a complex network analysis to understand the
hierarchical organization of functional brain networks, and we study how we can
explain the emergence of both small-world and modular features in the same network. We
capitalize on a well known dual-task paradigm, the psychological
refractory period, in which information from different sensory
modalities (visual and auditory) has to be coherently routed to
different motor effectors (in this experiment, the left and right
hand).

The combination of high-temporal resolution fMRI with novel network analysis tools
allows the study of the module properties and their synergy towards accomplishing a cognitive
task. A functional correlation network is derived from the fMRI phase
information. We implement percolation and scaling analysis methods
to uncover a highly modular functional operation and a network that is almost optimally
connected for efficient information flow.

\section{Material and Methods}

\subsection{Experimental design}

We use time-resolved fMRI \cite{menon1998mental}, based on analysis of
the phase signal \cite{sigman2007parsing}.
Time-resolved fMRI
is capable of identifying the
series of processing stages which unfold sequentially during the
execution of a compound dual-task \cite{dux2006isolation,sigman2008brain}.

The details of the experiments are described in \cite{sigman2007parsing},
and are briefly reviewed here.
Sixteen participants performed a dual-task paradigm: first a visual task of
comparing an Arabic numeral (target T1) to a fixed reference, with a
right-hand response and, second, an auditory task of judging the pitch
of an auditory tone (target T2) with a left-hand response. The
stimulus onset asynchrony (SOA) between T1 and T2 was varied between
0, 300, 900 and 1200 ms. In the course of this analysis we did not
detect significant differences in the resulting patterns of different
SOA conditions.

While subjects performed the dual-task, whole-brain fMRI
images were recorded at a sampling time (TR) of 1.5 s, and subsequently the phase and
amplitude of the hemodynamic response were computed \cite{sigman2007parsing}.
This activated map exhibits phases consistently
falling within the expected response latency for a task-induced
activation. As expected for an experiment involving visual
and auditory stimuli and bimanual responses, the responsive regions
included bilateral visual
occipito-temporal cortices, bilateral auditory cortices, motor,
premotor and cerebellar cortices, and a large-scale bilateral
parieto-frontal network \cite{sigman2008brain}. 
In this study we try to understand
the topology of the modular organization of this broad functional network during
dual task performance. For this purpose, we derived a large functional network of brain areas
by measuring the phase correlations in these responses for all pairs of voxels.
We then connected the highly-correlated pairs which gave us the brain cluster
network structure. 

\subsection{Phase correlations and functional brain network}

We use network theory concepts for the analysis of correlations
between different brain areas, based on the temporal activation of
these areas when a subject responds to external stimuli.  We
reconstruct the network topology of brain voxels, where a network link
indicates a high correlation in the phase-space activity of the two
connected voxels, and compare this structure with the corresponding
topology of the voxel location in the brain.

The time evolution of the phase of all brain voxels over 440 s was recorded
for each participant and each of the four SOA conditions, for a total of
64 measurements. For our analysis, we create a mask where we only
keep voxels which were activated in more than 75\% of the cases,
i.e. in at least 48 instances.

We want to detect the correlation between the phases of two voxels i
and j in the activated mask. The measure of correlation for vectors is
the co-directionality, i.e. we need to calculate the angle between the
two vectors. Therefore, the correlation $c_{ij}$ between two vectors
$\vec{r_i}$ and $\vec{r_j}$ is given, in general,
by $c_{ij}\equiv\vec{r_i}\cdot\vec{r_j}/|\vec{r_i}||\vec{r_j}|$, which
is equivalent to the cosine of the included angle, i.e. $c_{ij}=\cos(\theta)$,
where $\theta$ is now the phase difference $a_i-a_j$. We average
the correlation between any two voxels i and j in the activated mask
over roughly 40 trials of each experiment. The resulting correlation $p_{ij}$ between these two voxels is
then given by
\begin{equation}
p_{ij} = \frac{1}{N} \sum_{t=1}^N \cos\left[ a_i(t)-a_j(t) \right] ,
\end{equation}
where $N$ is the number of trials for a given combination of subject
and stimulus.  We link two voxels if their correlation is larger than
a threshold value $p$.  The resulting network is a representation of
functional relations among voxels for a specific subject and stimulus.

The topology of this network strongly depends on the value of $p$.
The variation of $p$ describes a percolation process. A large $p$
value enables isolated module identification, since only the strongest
(i.e. more correlated) functional links between voxels are
preserved. As $p$ is lowered, these modules get progressively merged
to larger entities and the emphasis is shifted towards large-scale
properties of the spanning network.

\begin{figure}
\centerline{
\includegraphics[height=7 cm]{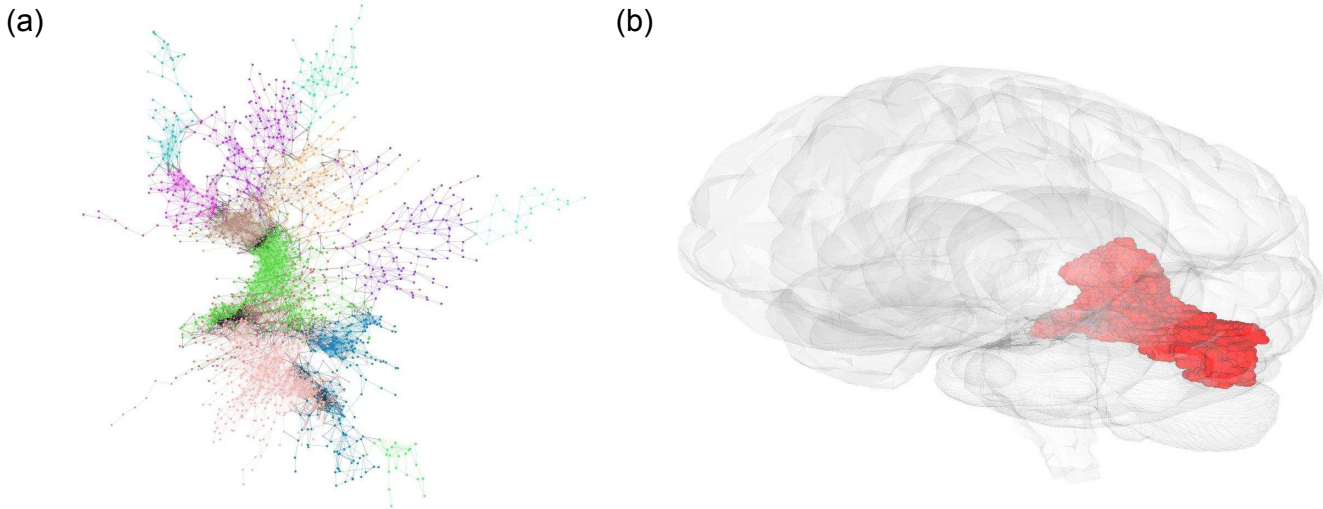}
}
\caption{
(a) Network representation of a brain cluster, as found by
the phase correlation between pairs of voxels.
(b) The same cluster in real-space representation, where each voxel
is now placed in its known location in the brain.
}
\label{FIGillustr}
\end{figure}

The complex network representation (Fig.~\ref{FIGillustr}a) reveals functional links between
brain areas, but cannot directly reveal spatial correlations. Since
voxels are embedded in space, we also study the topological features
of spatial clusters in three dimensions, where now voxels assume their
known positions in the brain and links between them are transferred
from the corresponding network (Fig.~\ref{FIGillustr}b), i.e. they are assigned according to
the degree of correlation between any two voxels, independently
of the voxels proximity in real space.

The above procedure yields a different network or spatial clusters for
each subject.  We study each of those networks and clusters separately
and show that they all carry statistically similar properties.  For
efficiency purposes, we focus our attention to the case of the largest
$p_c$ value where three clusters, including at least 1000 voxels,
emerge in each trial.  The spread of the corresponding $p_c$ values is
small, demonstrating a similar behavior in the brain response
of different subjects.

\subsection{Fractal analysis}

We analyze the resulting networks and the embedded three-dimensional
clusters in terms of their
fractal and modular properties. For the spatial representation, we
characterize the fractality of a connected cluster through the
standard Haussdorf dimension $d_f$. Starting from an arbitrary point
in a cluster, $d_f$ measures how the mass $N_f$ (number of voxels in
the same cluster) scales with the Euclidean distance $r$ from this
origin, i.e.:
\begin{equation}
N_f(r)\sim r^{d_f}.
\label{def_df}
\end{equation}
The exponent $d_f$ shows how
densely the area is covered by a specific cluster.

The box-covering technique is used for the fractal analysis of the
complex networks. 
A network (in our case each cluster) is first tiled with the minimum
possible number of boxes, $N_B$, of a given size $\ell_B$.
A box is defined as a union of nodes, all of which are at a distance from each
other smaller than a given threshold length, the box size $\ell_B$
(the distance between two nodes, $\ell$, is defined as the number of links along
the shortest path between those nodes in the functional brain network).

The fractality (self-similarity) of the network is quantified in the power-law relation
between the number of boxes needed to cover the network and the box size
$\ell_B$: 
\begin{equation}
N_B(\ell_B) = N_0 \ell_B^{-d_B}, 
\label{db}
\end{equation}
where $d_B$ is the fractal dimension (or box dimension) and $N_0$ is
the number of nodes in the original network
\cite{song05,goh06,song06,jskim07,radicchi08}.
Finite and small values of $d_B$ show that the
network has fractal features, where the covering boxes retain their
connectivity scheme under different scales, and larger-scale boxes
behave in a similar way as the original network.

\begin{figure}
\includegraphics[height=4 cm]{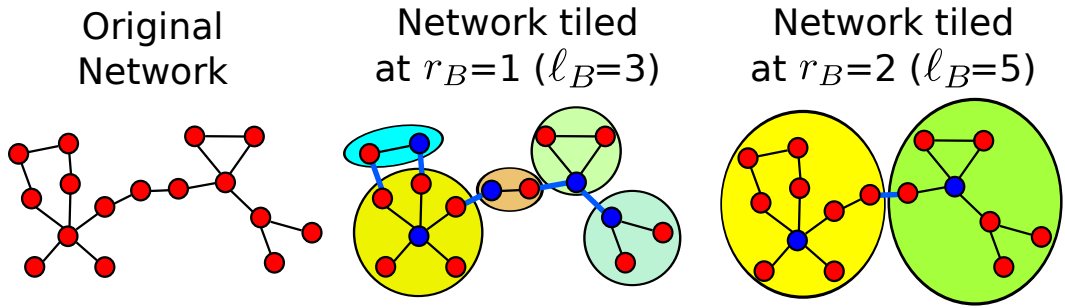}
\caption{ Demonstration of the MEMB box-covering algorithm.
For a given radius value, e.g. $r_B=1$ in the center panel and
$r_B=2$ in the right panel, we cover the network
with the smallest possible number of boxes. The diameter of the box, $\ell_B$,
(i.e. the distance between any two nodes in a box) is defined as
$\ell_B=2r_B+1$. First, we detect the smallest possible
number of box origins (shown with blue color) that provide the maximum
number of nodes (mass) in each box, according to an
optimization algorithm described in \cite{jstat}. Then, we build the boxes through
simultaneous burning from these center nodes, until the entire network
is covered with boxes.
For the calculation of modularity, we consider the boxes at each
$r_B$ value as separate modules. Then we calculate the ratio between the
number of links within the modules (black links) and the number of
links between modules (blue links).
}
\label{modules}
\end{figure}

The requirement that the number of boxes should be minimized poses
an optimization problem which can
be solved using a number of box-covering algorithms. The method that
we implement here is called Maximum Excluded Mass Burning algorithm (MEMB),
and the algorithm can be downloaded from
\url{http://lev.ccny.cuny.edu/~hmakse/soft_data.html}). The method
is roughly explained in
Fig.\ref{modules}.
The detection of modules or boxes in our work follows from the
application of this algorithm \cite{song05,jstat} at
different length scales.

The MEMB method starts by determining the minimum number of boxes
of radius $r_B$ required for a complete coverage of the network.
This radius is the distance from a box `center', so that by definition
all nodes in a box are within a distance from each other
smaller than $\ell_B=2 r_B+1$.
The method detects the nodes that will act as the centers of the boxes,
by calculating the mass around each node if it would act as a
center. The node with maximum mass around it is selected as a center
and we proceed iteratively to find the minimum number of such centers.
Once these nodes have been determined, the boxes are built by including
successive layers of nodes around the centers.
The details of the method are reported in \cite{jstat}.

The resulting boxes are characterized by the proximity between all
their nodes, at a given length scale and the
maximization of the mass associated with each module center.
Thus, MEMB detects boxes that also tend to maximize modularity. Different values of the box
diameter $\ell_B$ yield boxes of different size.  These boxes are then
identified as modules which at a smaller scale $\ell_B$ may be
separated, but merge into larger entities as we increase $\ell_B$.
Thus, we can study the hierarchical character
of modularity, i.e. modules of modules, and we can detect whether
modularity is a feature of the network that remains scale-invariant.

For this, we can extend the box-covering concept to act as a community detection
algorithm \cite{galvao}. MEMB identifies modules of size $\ell_B$, composed of highly
connected brain areas. Typical modularity approaches do not place
constraints on the size of the modules, but they focus on minimizing the
number of inter-module links. The MEMB approach, though,
has the additional advantage that modularity can be studied at different scales.
The requirement of minimal number of modules to
cover the network $(N_B)$ guarantees that the partition of the network
is such that each module contains the largest possible number of nodes
and links inside the module with the constraint that the modules
cannot exceed size $\ell_B$. 
This optimized tiling process gives rise to modules with the
fewest number of links connecting to other modules.
This implies that the degree of modularity for a given $\ell_B$ value
is maximized, and we can define a modularity measure, ${\cal M}$ through
\cite{girvan,amaral,caldarelli,lazaros}
\begin{equation}
{\cal M}(\ell) \equiv \frac{1}{N_B} \sum_{i=1}^{N_B} \frac{L_i^{\rm
    in}}{L_i^{\rm out}} .
\label{mo}
\end{equation}
Here $L_i^{\rm in}$ and $L_i^{\rm out}$ represent the
number of links that start in a given module $i$ and end either within
or outside $i$, respectively.  Large values of $\cal M$ (i.e. $L_i^{\rm
  out}\to 0$) correspond to a higher degree of modularity
\cite{lazaros}. 

The value of the modularity of the network $\cal M$
varies with $\ell_B$, so that we can detect the dependence of modularity
on different length scales, or equivalently how the modules themselves
are organized into larger modules that enhance the degree of
modularity. In the case that the dependence has a power law form,
we can define a modularity exponent $d_M$, through the relation:
\begin{equation}
{\cal M}(\ell_B) \sim \ell_B^{d_M} .
\label{modular}
\end{equation}

\section{Results}

\subsection{Percolation analysis reveals the modular structure}

We use percolation theory \cite{bunde-havlin} to identify the
functional clusters resulting from the correlation between the phases
of two voxels. The percolation problem is a paradigm of critical phase transitions
\cite{vicsek,stanley} which can be used to identify the functional clusters in the brain
network. In the simplest version of percolation, we can consider a lattice
where each bond is absent with probability $p$ or present with
probability $1-p$ ~\cite{bunde-havlin}. In lattices, it is well known
that there exists a critical probability $p_c$, below which the largest
cluster of connected bonds spans the whole length of the lattice, while
for $p>p_c$ only small isolated clusters survive.

In the case of the functional brain network, the corresponding probability
$p$ for the existence of a link between any two voxels in the brain is based
on the value of the phase correlation between them. For each
participant, we calculated the mass of the largest cluster as a
function of the percolation threshold $p$.  As explained above, in a
broad variety of systems in nature, the size of the largest cluster in
a percolation process remains very small and increases abruptly through a
phase transition, in which a single largest cluster spans the whole
system \cite{bunde-havlin}.  A single incipient cluster is expected to appear if the bonds in the network are
occupied at random without correlations, i.e. when the probability
to find an active bond is independent on the activity of all the other
bonds in the network. For the functional brain network our results revealed a more complex picture.

\begin{figure}[ht]
\centerline{ \resizebox{12.0cm}{!} { \includegraphics{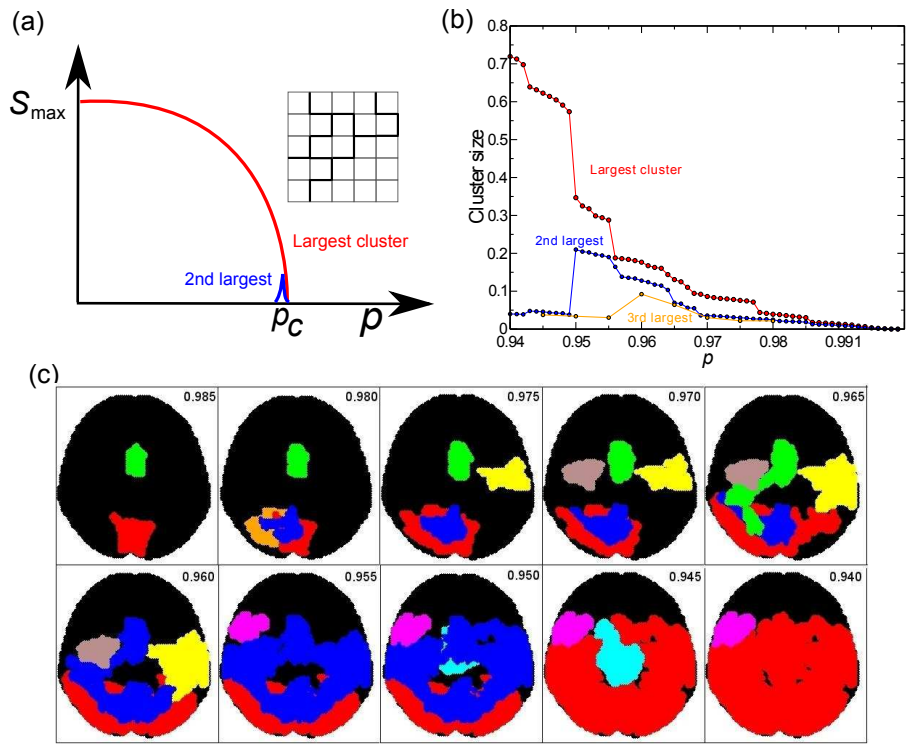}}}
  \caption{ 
  (a) Bond percolation in a 2-dimensional lattice. We remove a
    random bond with probability $p$, and with
    probability $1-p$ a bond remains (denoted by a solid black line).
    The lattice with solid black bonds is at the percolation
    transition $p=p_c$. (b) Evolution of modules at different
    thresholds.  The size of the three largest clusters as a function of the
  correlation threshold $p$ for a given subject. As we lower $p$ the
  cluster size increases in jumps, and new clusters emerge, grow,
  and finally get absorbed by the largest cluster.
  This behavior is significantly different from the random percolation in (a).
  (c) Cluster evolution. Brain clusters with more than 1000 voxels, as identified through
  correlation analysis for a given $p$ value.
}
\label{FIGjumps}
\end{figure}

We found that, for all participants in this study, the cluster size increased
progressively with a series of sharp jumps (Fig.~\ref{FIGjumps}) and not
with a single jump as expected for the simpler picture of uncorrelated
percolation. Moreover, in random percolation the second largest cluster
has a strong peak around $p_c$ and vanishes otherwise. In the brain network, the second
largest cluster also increases through jumps of absorbing smaller clusters.
Moreover, this second cluster remains comparable in size with the largest
cluster over a wider range of $p$. The evolution of these cluster sizes
with $p$ is a strong indication of strong correlations deviating from
a random process.

We identified each of the jumps in the largest cluster as a single percolation
transition focused on a region of the brain that is highly correlated
and therefore represents a well-defined module (Fig.~\ref{FIGjumps}).  These sharp
transitions are indicative of a marked modular structure in the
network. They indicate that at any given $p$ value there are many
isolated clusters in the brain network, which subsequently merge into
the largest cluster as $p$ decreases. This is a universal behavior
observed in all participants, and allows the
identification of functional modules, which we proceed to
study next.

The clusters identified by percolation analysis at a given threshold $p$
are functionally connected, but the nodes in such a cluster
are not necessarily clustered in space.
Thus, we first studied whether the percolation clusters had a consistent
spatial projection. The p-values at which clusters appear varied
across participants. To group the data, we measured, for each participant, the highest
correlation p-value for which there were at least three clusters of 1000 voxels each.  The
topography of these clusters reflected coherent patterns across
different individuals.  In virtually all participants we observed a cluster
covering the premotor, supplementary motor area (SMA) region, a
cluster covering the medial part of the posterior parietal cortex
(PPC) and a cluster covering the medial part of early retinotopic
cortex (area V1), along the calcarine fissure.

\begin{figure}[ht]
\centerline{ \resizebox{10.0cm}{!} { \includegraphics{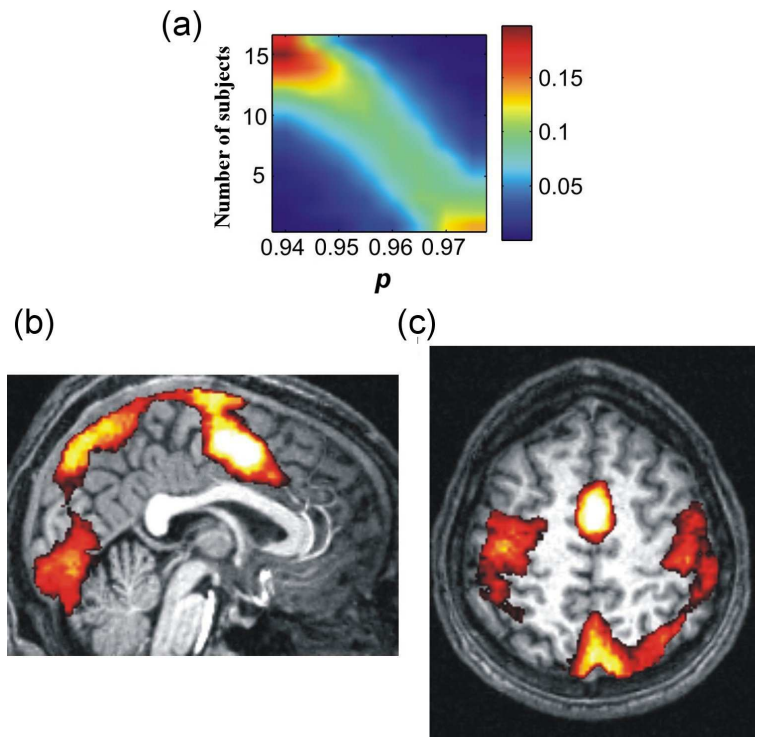}}}
  \caption{ 
The emerging clusters have consistent spatial projections. (a) 
The color denotes the fraction of the total number of voxels that appear to one of the three 
  largest clusters in $N$ subjects at a given percolation threshold
  $p$. As we reduce the threshold the peak shifts towards larger $N$
  values, i.e. the same voxels appear consistently in the largest clusters for all
  subjects. (b,c) Spatial distribution of the first
  percolation clusters (in subject counts).  The two brain slices show
  for the highest p-values the shared voxels. White bleached regions correspond to voxels which
  are included in the first percolation cluster for all subjects.  The
  SMA, a region involved in planning motor action is
  the only shared region for all subjects.}
\label{FIGproject}
\end{figure}

We then measured the likelihood that a voxel may appear in a
percolation cluster, by counting, for each voxel, the number of
individuals for which it was included in one of the first three
percolation clusters (Fig.~\ref{FIGproject}).

Clusters in the three main nodes, V1, SMA, PPC, are ubiquitously
present in percolation clusters and, to a lesser extent, voxels in the
motor cortex (along the central sulcus) slightly more predominantly on
the left hemisphere.

This analysis demonstrated that the correlation networks
obtained from each subject yielded percolation clusters with
consistent topographic projections. Next we focus on our main aim;
exploring the topology and scaling properties of the network modules
using fractal network analysis.

\subsection{Fractal analysis results}

For each of the 16 participants and each of the 4 SOA conditions
we calculated the resulting network through the phase
correlation.
Then, for each network, we estimated the 
percolation threshold that yields three clusters of at least 1000 voxels each. 
This results in a total of 192 clusters which were pooled together for the present analysis. 

\begin{figure}[ht]
\centerline{ \resizebox{16.0cm}{!} { \includegraphics{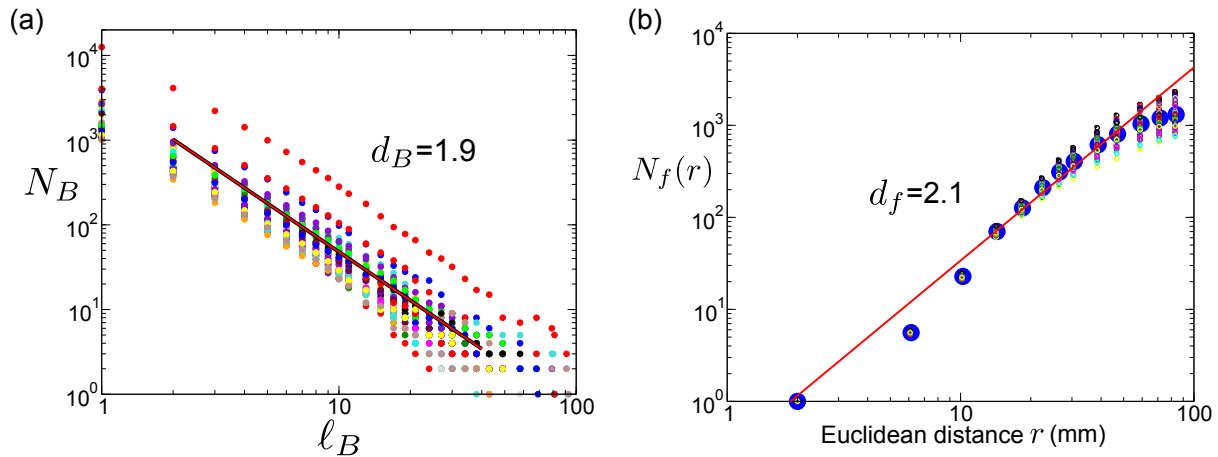}}}
  \caption{  (a) Fractal dimension $d_B$ of the network clusters.
The line is representative of the average dimension $d_B=1.9$.
  (b) Fractal dimension $d_f$ of the spatially embedded clusters.
The large points represent the number of nodes $N_f(r)$
included within a fixed distance $r$, averaged over all clusters, while
smaller points refer to individual clusters.
The fitted line corresponds to the average dimension $d_f=2.1$.
}
\label{FIGdb}
\end{figure}

We applied the box-covering algorithm \cite{song05,jstat}
to measure the fractal dimension $d_B$ of these 192 clusters.  
The fractal dimension $d_B$ was calculated separately for each cluster.
The resulting network fractal dimensions were distributed in a relatively
narrow range, with an average value $d_B=1.9 \pm 0.1$ (Fig.~\ref{FIGdb}a).

The cluster structure can be also probed by its topological features
when every node-voxel assumes its assigned location at the brain.
Each cluster identified by the box covering algorithm can be mapped to their anatomical projections,
where two voxels are still connected according to their correlation but their distance is now 
defined by the Euclidean three-dimensional spatial distance $r$  (Fig.~\ref{FIGdb}b).
This mapping allows the use of the classical fractal dimension in real space
for the study of the structure of these functional clusters in the brain.

The method that we use to calculate the fractal dimension here is an alternative
method to the one used in Ref.~\cite{newpnas}. There, $d_f$ was calculated by measuring
the number of nodes, $N_C$, in a cluster as a function of the cluster diameter. Here, for every
cluster we start from a random point and open a circle of radius $r$ and measure the number
of nodes $N_f(r)$ in this circle. The dependence of $N_f(r)$ on $r$ for this cluster gives its
fractal dimension, and the process is repeated for all clusters.
The scaling of the mass $N_f(r)$ (i.e. number of nodes in the cluster) included in a sphere with
Euclidean radius $r$ follows the power-law form of Eq.~(\ref{def_df}).
The calculation of the individual
Euclidean fractal dimensions yields an average of $d_f=2.1\pm0.1$
(Fig.~\ref{FIGdb}b), which is similar for all clusters, and which is exactly the same as
the one found in \cite{newpnas}.
The network fractal dimension of all clusters was systematically lower than the real-space
fractal dimension, which was in the range 2-2.4.

\begin{figure}[ht]
\centerline{ \resizebox{16.0cm}{!} { \includegraphics{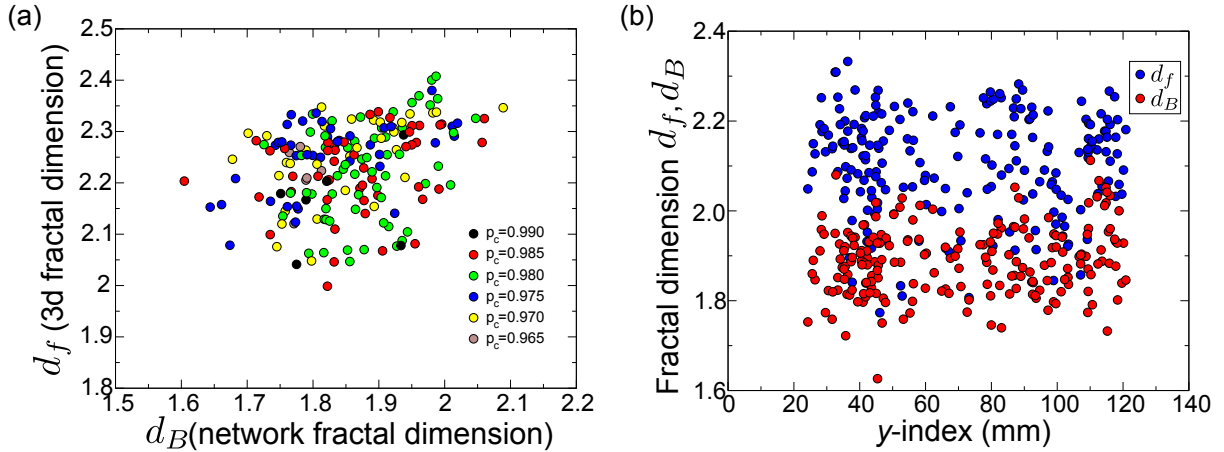}}}
  \caption{  Consistency of the fractal dimension calculations. (a) Cross-plot of $d_f$
vs $d_B$ for individual brain clusters. The colors correspond to the threshold values 
of $p_c$ where the first percolation transition was identified. (b) The network fractal
dimension, $d_f$ (blue), and the three-dimensional fractal dimension, $d_B$ (red) as a function
of the location of each cluster. This location corresponds to the center of mass, and
is expressed through the $y$-index, posterior to anterior.
}
\label{FIGdfdb}
\end{figure}

It is possible that the difference between the fractal dimensions of individual clusters can be due to
systematic variations, influenced by various factors. We performed a number of tests
to identify the stability of these calculations.
In Fig.~\ref{FIGdfdb}a we show a cross-plot for the exponents $d_B$ and $d_f$, as calculated
for each individual cluster. The value of $d_B$ was systematically below $d_f$. 
From the same plot we deduce that the value of the percolation transition does not influence
the fractal dimension, since the different $p_c$ values of different clusters yield
a uniform spreading of the fractal dimensions.
It is also possible that the location of the brain clusters may have an effect on their
fractal character. Our results do not provide any evidence towards this direction, either.
In Fig.~\ref{FIGdfdb}b we plot the exponents $d_B$ and $d_f$ for each cluster as a function
of the $y$-coordinate of the cluster's center of mass, i.e. increasing $y$ indexes corresponds
to moving from the posterior to the anterior part of the brain. It is obvious that there is no
systematic variation of the exponents in different locations.
The above results emphasize the robustness of the fractal structure and indicate that we
can consider the averages over all those structures to be representative of a typical brain
module.

We can now characterize each single cluster, both at the functional level
and at the topological level (i.e. the shape that the cluster assumes in the brain).
Together, these results indicate that none of the clusters fill the 3D space 
densely; although the objects are embedded in three-dimensions
their fractal dimension $d_f$ is significantly smaller than 3.
The network structure provides information on functional clusters, since
it relates areas that are highly correlated independently of their physical proximity.
Since the network fractal dimension $d_B$ is even smaller than $d_f$, 
connections are fewer than one would expect through nearest-neighbor connections
only. In simpler words, clusters do not form densely connected
neighborhoods.

\subsection{Modular structure}

\begin{figure}[ht]
\centerline{ \resizebox{8.0cm}{!} { \includegraphics{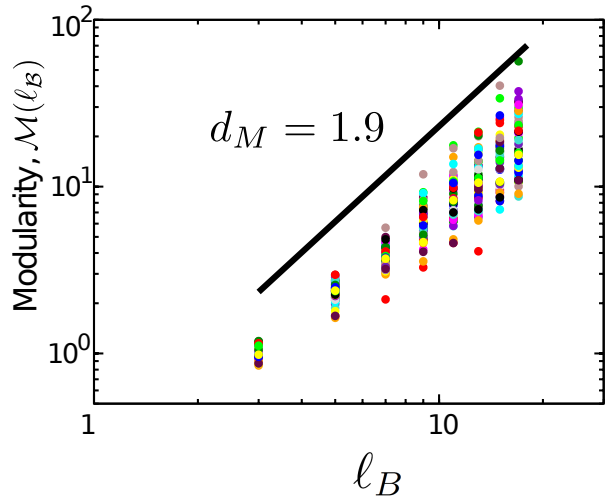}}}
  \caption{  Modularity as a function of $\ell_B$ for different clusters.
The average value of the exponent is $d_M=1.9$, shown by the solid line.
}
\label{FIGmodul}
\end{figure}

In the Materials and Methods section we described how we can use
the optimal MEMB coverage of the network with $N_B$ nodes for
a given $\ell_B$ value, in order to characterize the network modularity.
Analysis of the modularity Eq.~(\ref{mo})
in Fig. \ref{FIGmodul} reveals a monotonic increase of ${\cal M}(\ell_B)$
with a lack of a characteristic value of $\ell_B$.  Indeed, the data can
be approximately fitted with a power-law functional form, Eq.~(\ref{modular}),
which is characterized by the modularity exponent $d_M$.  We
analyze the resulting networks of different subjects and we find
that $d_M = 1.9\pm0.1$ is approximately constant over different individuals.
 (Fig. \ref{FIGmodul}).  
 
This value reveals a considerable degree of modularity in the entire system
as evidenced by the network structure.
For comparison, a random network has $d_M=0$ and a uniform lattice has $d_M=1$ \cite{lazaros}).
The lack of
a characteristic length-scale in the modularity shown in
Fig. \ref{FIGmodul}
suggests that the modules appear at all length-scales, i.e. modules
are organized within larger modules in a self-similar way, so that the
inter-connections between those clusters repeat the basic modular
character of the entire brain network.  Thus, the modular organization of the network remains
statistically invariant when observed at different scales.


\subsection{Short-cut wiring is optimized for efficient flow}

A major advantage of the present analysis approach is that the analysis
of the type of short-cuts present in the brain networks can convey a notion of 
optimal navigability in the network.



The addition of long-range links can turn the balance of a network structure
towards either a self-similar structure with significant modularity but poor
transfer or towards a small-world structure with very efficient flow at
the cost of modularity (specialization). A small number of such shortcuts,
quantified through renormalization group analysis \cite{rozenfeld},
has been shown to provide the optimal trade-off between these two properties.
In the case of the brain clusters the need for specialization/modularity is obvious,
as also shown in the previous section,
so it is important to understand how shortcuts influence the efficiency of signal transport
in these structures.

In order to study how the modules that we recovered by the first percolation transition
integrate at a larger scale, we also considered another percolation transition
that corresponds to the emergence of a spanning cluster. We chose this transition
as the correlation point where the largest cluster is equal to half of the total
size. This global network connects practically all the smaller brain modules.

\begin{figure}[ht]
\centerline{ \resizebox{16.0cm}{!} { \includegraphics{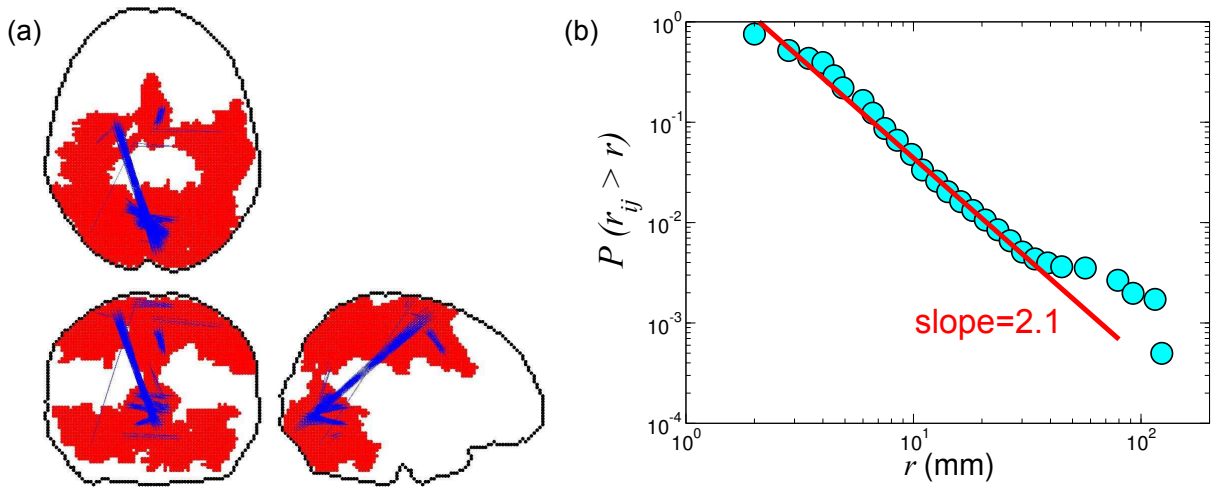}}}
  \caption{ 
(a) Real-space representation of the network at the second global percolation.
The largest component is half the total mass. The blue lines highlight the
longest links in Euclidean distance, which correspond to the weak ties.
(b) Cumulative
probability distribution $P(r_{ij}>r)$ of Euclidean distances $r_{ij}$
between any two voxels that are directly connected in the correlation
network. The straight line fitting yields an exponent
$\alpha-1=2.1\pm0.1$ indicating optimal information transfer with
wiring cost minimization.
}
\label{FIGPr}
\end{figure}

We probed the connectivity for this network, by analyzing
the distance distribution of the links in the network,
i.e. the Euclidean distance between any two voxels that are connected through their
phase correlation (Fig.~\ref{FIGPr}a).
We find an approximately power law distribution (Fig.~\ref{FIGPr}b) of the form:
\begin{equation}
P(r) \sim r^{-\alpha},
\end{equation}
with a short-cut exponent $\alpha \approx 3.1$. 
The value of this exponent is very significant, since 
it approximately satisfies the scaling relation with the fractal dimension of the brain network:
\begin{equation}
\alpha = d_f + 1.
\end{equation}
Such a scaling relation was recently \cite{li} found to optimize the transfer of
information across a network with fractal dimension $d_f$ when the
short-cuts in the network are added with a cost constraining the
number of total links. Thus, our scaling and modular analysis suggests that,
taking into account the spatial restrictions, the functional
behavior of the brain is optimally wired for facilitating efficient information
transfer among different areas. 

\section{Discussion}

Our analysis revealed a fractal structure for the individual brain clusters.
These clusters have a consistent topological behavior and are located
at the areas that correspond to the expected brain responses.
These modular stuctures present consistent fractal properties,
both at the functional level and at a topological level. This indicates
that the individual processing units that we recover do not have
significant small-world properties. In constrast, when we include
weaker correlations, the modules that appear at smaller scales are connected
through long-range links. These shortcuts give a small-world
character to the brain network as a whole, i.e. when studied at
scales larger than an individual module.

The study of the distribution for these links suggests interestingly that they
are optimizing transfer network properties, by also considering
the wiring cost.
In simpler terms, this topology does not minimize
the global connectivity, simply to connect all the nodes; instead
it minimizes the amount of wire required to achieve the goal
of shrinking the network to a small-world.

The existence of modular organization of strong ties in
a sea of weak ties is reminiscent of the structure found
to bind dissimilar communities in social networks. Granovetter's
work in social sciences \cite{granovetter} proposes the
existence of weak ties to cohese well-defined social groups
into a large-scale social network. Such a two-scale structure
has a large impact on the diffusion and influence
of information across the entire social structure. Our observation
of this two-layer organization in brain networks
suggests that it may be a ubiquitous natural solution to
the puzzle of information flow in highly modular structures.

Previous studies have found that wiring of neuronal
networks at the cellular level is close to optimal \cite{song}.
Specifically it is found that long-range connections do
not minimize wiring but achieve network benefits. In
agreement with this observation, at the mesoscopic scale
explored here, we find an optimization which reduces
wiring cost while maintaining network proximity. An intriguing
element of our observation is that this minimization
assumes that broadcasting and routing information
are known to each node. How this may be achieved -
what aspects of the neural code convey its own routing
information - remains an open question in Neuroscience.

\begin{acknowledgments}
We thank D. Bansal, S. Dehaene, S. Havlin, and
H.D. Rozenfeld for valuable discussions. L.K.G. and H.A.M. thank the
NSF-0827508 Emerging Frontiers Program for financial support. M.S. is supported
by a Human Frontiers Science Program Fellowship.
\end{acknowledgments}

\end{document}